# A Multi-stage Transfer Learning Framework for Diabetic Retinopathy Grading on Small Data

Lei Shi[1,2], Bin Wang[3] and Junxing Zhang[1,*], *IEEE Member*

[1] College of Computer Science, Inner Mongolia University, Hohhot, China.
[2] Baotou Medical College, Baotou, China.
[3] Department of Ophthalmology, The First Affiliated Hospital of Baotou Medical College, Baotou, China.
Email: shilei@mail.imu.edu.cn; junxing@imu.edu.cn

*Abstract*—Diabetic retinopathy (DR) is one of the major blindness-causing diseases currently known. Automatic grading of DR using deep learning methods not only speeds up the diagnosis of the disease but also reduces the rate of misdiagnosis. However, problems such as insufficient samples and imbalanced class distribution in small DR datasets have constrained the improvement of grading performance. In this paper, we apply the idea of multi-stage transfer learning into the grading task of DR. The new transfer learning technique utilizes multiple datasets with different scales to enable the model to learn more feature representation information. Meanwhile, to cope with the imbalanced problem of small DR datasets, we present a class-balanced loss function in our work and adopt a simple and easy-to-implement training method for it. The experimental results on IDRiD dataset show that our method can effectively improve the grading performance on small data, obtaining scores of 0.7961 and 0.8763 in terms of accuracy and quadratic weighted kappa, respectively. Our method also outperforms several state-of-the-art methods.

## I. Introduction

Diabetic Retinopathy (DR) is one of the main complications of diabetic patients. It can easily lead to blindness, if not discovered and intervened at an early stage. According to current international standards, DR can be graded into five levels or classes: normal, mild, moderate, severe and proliferative DR, corresponding to grades DR0 to DR4 [1, 2].

At present, the main method for diagnosing the severity of diabetic retinopathy relies on fundus images and diagnosis results given by ophthalmologists. However, this process will take a long time for patients to wait for, and diagnosis results are often influenced by the number and experience of ophthalmologists. Therefore, it is necessary to develop a computer-aided diagnosis system to help ophthalmologists make correct decisions and speed up the overall process of DR diagnosis.

In recent years, medical image diagnosis technology based on deep learning has made great progress. Particularly for eye disease grading, researchers have already proposed many valuable methods [3-8]. However, problems and challenges remain in the difficult task of DR grading on small datasets: (1) small datasets only contain 100-level images in total or even less than that, which makes the deep model difficult to learn. Transfer learning is often used to solve this problem. However, existing transfer learning studies only consider transferring features from one single large dataset directly to the target small datasets. However, this transfer strategy does not make full use of other related DR datasets; (2) The class distribution is imbalanced almost in all DR datasets, especially in small ones. For instance, in IDRiD training set [9], DR0 and DR2 account for 32 % and 33%, respectively. However, the most serious DR4 only accounts for 12% of the total data volume, as shown in Fig. 1. The imbalance greatly increases the possibility of misclassification of the low sample data, and will ultimately weaken generalization ability of the grading model. The aforementioned problems make it difficult to effectively enhance the grading performance of small DR datasets. Current studies have not proposed simple and effective solutions to these problems.

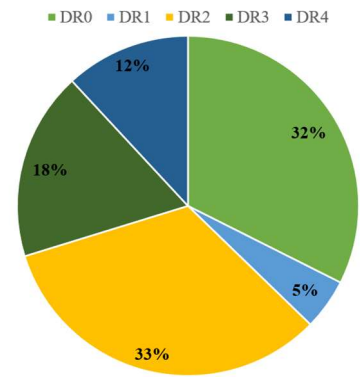

Fig. 1. Imbalanced class distribution of IDRiD training set.

For the first problem, this paper applies the idea of multi-stage transfer learning, which enables the grading model of small dataset to fully learn feature representation of the large-scale and medium-scale datasets. For the second problem, based on the concept of "decoupling representation learning and classifier learning" proposed in the literature [10][11], our work directly learns with imbalanced data at the stage of feature representation learning. At the stage of classifier learning, based on the best weight of the feature representation learning, we freeze all layers except for the last fully connected layer. The last layer is reinitialized and trained again. Here we introduce a class-balanced loss [12], initially for natural image classification, into our task as the loss function. Training for only a few epochs can be effective for classifier learning. Experimental results show that compared to state-of-the-art methods, the proposed method improves both accuracy and quadratic weighted kappa effectively.

---

* Corresponding author: Junxing Zhang (junxing@imu.edu.cn).

The main contributions of this paper are as follows:

- As far as we know, this paper applies the idea of "multi-stage transfer learning" into the DR grading task for the first time, so that grading performance can be further improved on the small dataset based on the "knowledge" of large and medium-scale datasets. This method also provides useful thoughts for other learning tasks on small data.
- A simple training method for class-balanced loss function is proposed in the DR grading task. Instead of training with class-balanced loss in an end-to-end manner, we decouple the entire training process into the feature representation learning stage and classifier learning stage. The standard cross-entropy loss is used for training in the feature representation learning stage, while the class-balanced loss is used for training in the classifier learning stage. Without using complicated hyperparameter tuning strategies such as learning rate decay, warmup, etc. [12], grading performance can be further improved by training the network only for a few epochs in the classifier learning stage. Classifier learning is not isolated and it is based on feature representation learning. The above two stages together serve the role of class-balanced loss used in this paper. This method can reduce the difficulty of training class-balanced loss function while mitigating data imbalance of different classes.

## II. RELATED WORK

Many works have been published in the area of automated diagnosis on DR and diabetic macular edema (DME) with deep learning. Li et al.[3] proposed and published a large-scale DR dataset named DDR, which provided useful data for evaluating newly proposed algorithms and exploring clinical applications. For dealing with the problem of lacking large balanced training data, Zhou et al. [4] proposed a so-called DR-GAN to synthesize high-resolution fundus images for effective data augmentation. The same team also proposed a collaborative learning method [5] with attention mechanism and adversarial architecture to jointly improve the performance of disease grading and lesion segmentation. Li et al. [6] paid attention to the joint classification problem of DR and DME and proposed a cross-disease attention network. However, it has not been noticed in the above methods that use of different scales of DR datasets may have a positive impact on the grading performance of small datasets.

There are relatively few researches focused on class imbalance of DR datasets. He et al. [8] proposed a category attention module (CAB) to alleviate the problem of imbalanced data, which is able to learn the characteristics of a specific grade and enlarge the distance of different grades. However, compared to class-balanced loss, CAB failed to solve the imbalance problem of DR grading intuitively and effectively although the grading performance can be improved by adding CAB into the network. In terms of solving the misclassification caused by DR class correlation, Galdran et al. [13] added a cost-sensitive regularization to three conventional loss functions to improve the DR grading performance. Liu et al. [14] proposed a graph convolutional network with an adjacency matrix, which introduced prior knowledge of class dependency to improve grading performance.

## III. PROPOSED METHODS

### A. Datasets

Because the datasets are used in the description of our method, we first introduce three datasets used in this work at the beginning of this section.

**Eyepacs**[15]: The 2015 Kaggle official competition dataset, a total of 88702 fundus images, including 35,126 in training set and 53,576 in testing set. DR is graded into five classes consistent with international protocols. Though Eyepacs has a large amount of data, there exists much noise in both training set and testing set. Researchers claimed that many ungradable images are classified as DR0 and there are some poor-quality images in them.

**DDR** [3]: A fundus dataset published by Nankai University in 2019, which is collected from Chinese patients. The DR grading subset consists of 13,673 fundus images with pixel-level annotations, including 6835 training images, 4105 testing images, and 2733 valid images. These images are with six classes, classifying low-quality or ungradable images as DR5 except for the first five standard classes. In our experiments, in order to be consistent with current international protocols and other datasets, the sixth-class images are removed.

**IDRiD** [9]: The 2018 ISBI diabetic retinopathy challenge dataset, which is collected from Indian patients. There are a total of 516 pictures used for disease grading, including 413 in the training set and 103 in the testing set. This dataset is characterized with small data, but high quality and accurate labeling. This dataset is used as the target small dataset to evaluate the grading performance of different methods. Examples with different severity of DR from this dataset are shown in Fig. 2.

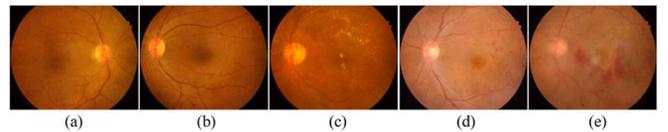

Fig. 2. Examples of fundus images with different severity of DR:(a) normal; (b) mild; (c) moderate; (d) severe; (e) proliferative.

### B. Feature Representation Learning

In this paper, Efficientnet-b5[16] is adopted to be the backbone network. Efficientnets are powerful models for classification tasks proposed by Google in 2019. Considering the high resolution of fundus images, it may improve grading performance to use larger input resolution. Therefore, we start our experiment from Efficientnet-b0 and finally enlarge the network to Efficientnet-b7. We find that networks larger than Efficientnet-b5 will not further improve the grading performance.

In the feature representation learning stage, the idea of multi-stage transfer learning is used to learn continuously from three different DR datasets. Thus, the final small dataset can learn feature representation from the former large and medium-scale datasets. The overall framework proposed in this paper is illustrated in Fig.3.

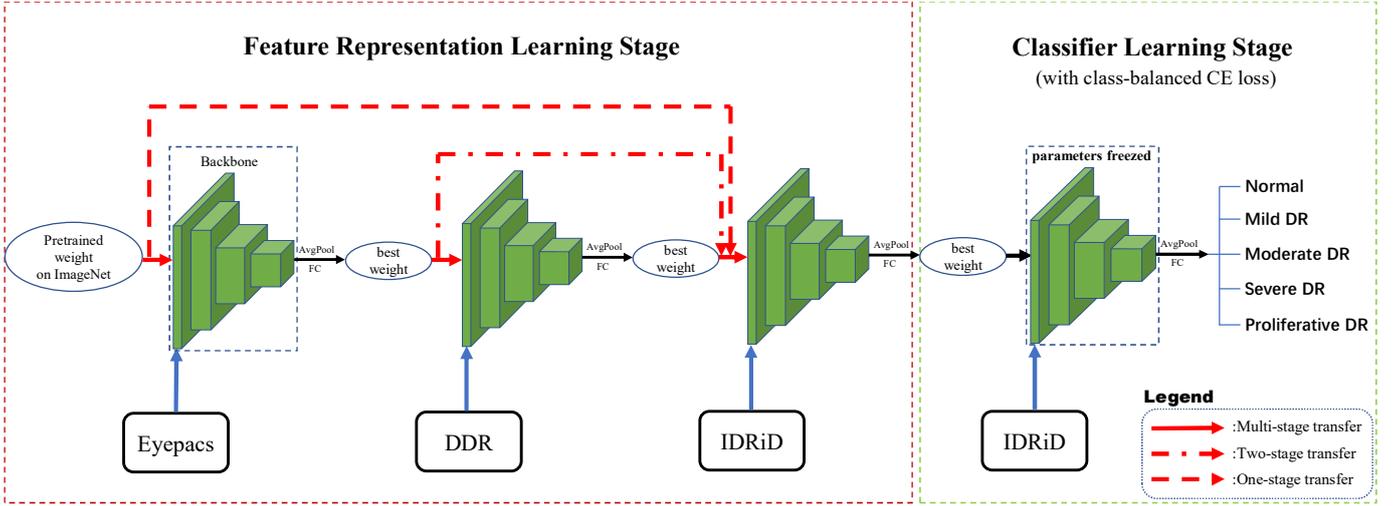

Fig. 3. The framework of our proposed scheme includes the feature representation learning stage and classifier learning stage. In the feature representation stage, the model learns the training sets of ImageNet, Eyepacs, DDR, and IDRiD in a multi-stage transfer way, and then evaluated on the IDRiD test set. For comparison with multi-stage transfer in the ablation studies followed up, paths of one-stage transfer and two-stage transfer are also shown with dashed lines. One-stage transfer refers to transferring directly from ImageNet and two-stage transfer refers to transferring first from ImageNet and then Eyepacs to IDRiD. In the classifier learning stage, based on the weight obtained in the feature representation stage, the fully connected (FC) layer is reinitialized and the classifier is trained again with the training set of IDRiD. A class-balanced cross-entropy loss is used to alleviate the imbalance problem, after which the model is evaluated on the IDRiD test set.

As shown in Fig. 3, our proposed scheme is structured with a feature representation learning stage and a classifier learning stage. The main task of the feature representation stage is to fully learn the features of different datasets with the idea of multi-stage transfer learning, while the main task of the classifier learning stage is to retrain the classifier layer with class-balanced loss to ease the class-imbalance problem, and then improve the prediction ability of the model for severe diseases (DR3 and DR4).

In the feature representation learning stage, the model learns the training sets of ImageNet [17], Eyepacs, DDR, and IDRiD serially in a multi-stage transfer path (shown with the solid line in Fig. 3), using the pretrained best weight based on the previous dataset. More specifically, we carry on the first step on Eyepacs based on ImageNet pretrained weight and save the best weight (with the lowest validation loss) for the next step; then the second step learning on DDR based on Eyepacs pretrained best weight. The first two steps of multi-stage transfer learning are both trained with smaller epochs; thus, the models can learn the representation of different datasets without overfitting on a certain one or damaging ability to learn new tasks. The third step is to carry on transfer learning on IDRiD based on DDR pretrained best weight and this step is trained with a larger epoch. The trained model is evaluated on the IDRiD test set.

### C. Classifier learning

For the imbalance problem of the small datasets, we adopt the re-weighting method to solve. To be specific, we assign weights to each of the five classes of DR based on standard cross-entropy loss: large weights for minority classes, and small weights for majority classes. Assuming that the prediction output tensor of the model is $p = [p_0, p_1, ..., p_{C-1}]$, where C is the number of classes, and the corresponding ground truth is y. Thus, the standard cross-entropy loss (CE loss) can be calculated as:

$$\text{CEloss}(p, y) = -\log\left(\frac{\exp(p_y)}{\sum_{j=0}^{C-1} \exp(p_j)}\right) \quad (1)$$

There are many ways to design weights, such as directly taking inverse class frequency [18]. In this paper, we adopt a class-balanced cross-entropy loss (CBCE loss) proposed in the literature [12], namely:

$$\text{CBCEloss}(p, y) = \text{weight}[y] \cdot \text{CEloss}(p, y)$$
$$= -\frac{1-\beta}{1-\beta^{n_y}} \log\left(\frac{\exp(p_y)}{\sum_{j=0}^{C-1} \exp(p_j)}\right) \quad (2)$$

The hyperparameter β ranges in [0,1) in class-balanced loss so that the weight is between no re-weighting and re-weighting by inverse class frequency. Here we set β to be 0.9999. $n_y$ is the number of corresponding grade in IDRiD training set, and the weight tensor can be calculated according to the above values. However, it often requires lots of tuning skills to directly train with this class-balanced loss on original DR data from end to end. According to our multiple experiments, we find that directly applying the implementation strategy in literature [12] will hurt the performance rather than improve it.

In order to avoid the uncertainty caused by overuse of hyperparameter tuning skills, inspired by literature [10] [11], we decouple feature representation learning and classifier learning, as shown in Fig.3. The feature representation stage adopts standard (unweighted) cross-entropy loss to learn from the original data distribution of the three datasets. No re-balancing strategies are added except for data augmentation to fully learn the feature representation information. Based on the best weight of the feature representation learning stage, we freeze all layers, reinitialize the fully connected (FC) layer randomly, and then only retrain this FC layer on the training set of IDRiD. We use the CBCE loss and a larger learning rate in the training process. After no more than 5 epochs, a significant performance improvement can be seen when evaluating on the test set.

## IV. EXPERIMENTS AND RESULTS

### A. Evaluation Metrics

To quantitively evaluate the effectiveness of methods proposed in this paper, we introduce overall accuracy (abbreviated as Acc below) and quadratic weighted kappa (abbreviated as Kappa below), which are widely used in papers and competitions. Acc reflects the overall prediction accuracy, namely the ratio of the number of images that are correctly predicted to the total number of images in the whole training/test set.

For better reflecting the grading performance on imbalanced dataset and measuring the agreement between ground truths and predictions, Kappa is also introduced to evaluate the performance of different methods. This metric varies from 0 to 1. The higher the value, the better the grading performance, especially for the class-imbalanced case.

### B. Implementation Details

Our backbone network uses Efficientnet-b5 with an input resolution of 456*456. In order to increase the data diversity, we use random horizontal flip, random vertical flip, random rotation, and color jitter as data augmentation. For IDRiD, we use stratified sampling and select 10% of the data in the training set as the validation set. For the feature representation stage training of DDR and IDRiD, batchsize is set to be 8, the learning rate is set to be 0.001, the loss function is standard cross-entropy loss, the optimizer is SGD, and the momentum is set to 0.9. For the value of epoch, Eyepacs is set to be 30, DDR is set to be 18, and IDRiD is set to be 150. For the classifier learning stage training, the learning rate is set to be 0.01, the loss function adopts the CBCE loss, and the epoch is set to be 5. We find that increasing the epoch will not further improve the grading performance in the classifier learning stage after a lot of experiments.

For each backbone output in Fig.3, we select the best weight (with lowest validation loss) as the pretrained weight of next step. We use Pytorch as the experiment framework. We use an NVIDIA A100 for our experiments.

### C. Ablation Studies

TABLE I. THE ABLATION STUDY WITH DIFFERENT SCHEMES ON THE IDRiD DATASET.

| Backbone | Method | Acc | Kappa |
|---|---|---|---|
| Efficientnet-b5 | One-stage transfer | 0.5631 | 0.6436 |
| | Two-stage transfer | 0.7476 | 0.8304 |
| | Two-stage transfer + CBCE loss | 0.7476 | 0.8670 |
| | Multi-stage transfer | 0.7573 | 0.8316 |
| | Multi-stage transfer + CBCE loss | **0.7961** | **0.8763** |

Table I shows the results of ablation experiments. On the IDRiD dataset, compared to the two-stage transfer learning scheme, the scheme adding class-balanced cross-entropy loss (CBCE loss) can increase Kappa by 3.66% and the scheme of multi-stage transfer learning (MSTL) can increase Acc by 0.97% and Kappa by 0.12%. After using both multi-stage transfer learning and CBCE loss, Acc and Kappa further increase 3.88% and 4.47%, respectively. We also investigated the parallel transfer scheme, which means parallel training with multiple datasets and using the ensemble weight to grade small data. We use two ways for weights ensemble: sum and average. We find that the performance of either ensemble way is weaker than any single weight as well as the multi-stage transfer method. The experiment results show that our methods are effective for classification task of small datasets.

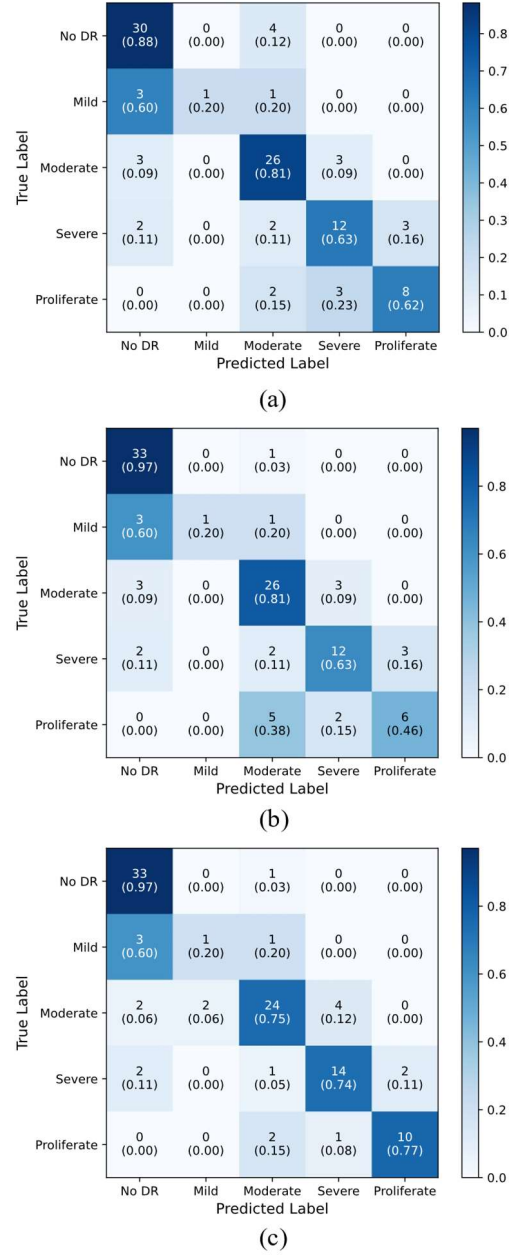

Fig. 4. Confusion matrix of different methods: (a) Two-stage transfer; (b) Multi-stage transfer; (c) Multi-stage transfer + CBCE loss. The value in the bracket refers to normalized number over each row.

Fig.4 shows the confusion matrix corresponding to different methods. The value of Acc, which is different from class average accuracy, can be calculated as the sum of the main diagonal elements of the confusion matrix divided by the sum of all the elements of the matrix. After introducing multi-stage transfer and CBCE loss, Acc is steadily improving from Fig.4(a) to

Fig.4(c). However, relying only on Acc is difficult to fully reflect the consistency between the prediction results and the true labels. In order to better show the difference between the predicted distribution and the true distribution, we need to analyze Kappa further. If the predicted labels are completely consistent with the true labels, all elements in the matrix should be distributed on the main diagonal, Kappa should be 1. Therefore, it gives a better-predicted result that data distribution of the confusion matrix gets closer to the main diagonal (Kappa gets closer to 1). The Kappa value of Fig.4(a) to Fig.4(c) has been increasing steadily and the method of adding both multi-stage transfer learning and CBCE loss has the closest consistency to the ground truth. Because distance weight is set in Kappa, for the false prediction, the farther the distance from the true label, the greater penalty will be given. In other words, for each grade, the more predicted label close to the main diagonal value of certain row, the more helpful to the improvement of Kappa.

Comparing Fig.4(b) and Fig.4(c), grading performance of the serious sample classes is significantly improved. For grade 3, the number of correct predicted images increases from 12 to 14 after adding CBCE loss; For grade 4, not only the number of correct predicted images increases by 30% but also the data distribution is more concentrated to the main diagonal position. In general, the introduction of multi-stage transfer learning and CBCE loss has contributed to the improvement of Acc and Kappa, respectively and jointly.

From the confusion matrix, false positive and false negative rates can also be calculated for a certain class. For example, in Fig.4(c), if we consider DR4 as the positive class, we can deem DR0 to DR3 as the negative class, thus converting the multi-class grading problem into a binary grading problem. It can be easily obtained that FP is 2 and FN is 3, the false positive rate is 2/90 and the false negative rate is 3/13. Both rates are the lowest in all three methods in Fig.4. Similarly, for Fig.4(c), both the false positive rate and the false negative rate of DR3 are also the lowest. Our methods are effective for serious classes, which are paid more attention clinically.

### D. Comparisons with State-of-the-art Methods

Table Ⅱ shows the comparison between our method and the top three places on the DR grading leaderboard of IDRiD Sub-challenge-2 [19]. As seen from this table, Acc obtained by our method exceeds that of LzyUNCC by 4.85%. It is worth noting that the input resolution of our proposed scheme is much lower than that of LzyUNCC (456*456 vs. 896*896). Meanwhile, in order to verify the effectiveness of our method further, we compare it with three state-of-the-art DR grading methods. CF-DRNet [20] firstly divides the input images into two categories through the coarse-grained network: with DR and without DR, and further divides the images with DR into grades 1 to 4 through a fine-grained network. LA-NSVM [21] proposes lesion-aware attention with neural support vector machine for DR grading. DR|GRADUATE [22] improves the results of DR grading by outputting a grade uncertainty. Compared with CF-DRNet and LA-NSVM, the Acc of our method exceeds 19.41% and 16.37%, respectively. Compared with DR|GRADUATE, the Kappa of our method exceeds 3.63%. In summary, the grading performance of our proposed method is obviously competitive.

TABLE II. GRADING PERFORMANCE COMPARISON WITH STATE-OF-THE-ART METHODS ON THE IDRiD DATASET.

| Methods | Acc | Kappa |
|---|---|---|
| Mammoth [19] *rank3* | 0.5437 | - |
| VRT [19] *rank2* | 0.5922 | - |
| LzyUNCC [19] *rank1* | 0.7476 | - |
| CF-DRNet [20] | 0.6020 | - |
| LA-NSVM [21] | 0.6324 | - |
| DR|GRADUATE [22] | - | 0.84 |
| Ours | **0.7961** | **0.8763** |

### E. Qualitative Analysis and Interpretability

To further illustrate effectiveness of our proposed methods, Gradient-weighted Class Activation Mapping (Grad-CAM) [23] is introduced to produce visual explanations for prediction. Four randomly selected fundus images and their corresponding class activation maps are shown in Fig.5. These fundus images are all labeled as proliferative DR, namely the most serious class to be deeply concerned. As can be seen from left column, the fundus of proliferative DR patients is usually accompanied by severe preretinal or subretinal hemorrhages, as indicated in green boxes. The activation maps with our method can effectively focus on the hemorrhagic areas in all four images. The visualization results are consistent with the ophthalmologist's experience to diagnosis DR. In other words, ophthalmologists also pay attention to the activated red areas. The visualization results demonstrate that the proposed method is effective and can provide interpretability for automatic DR grading in clinical applications.

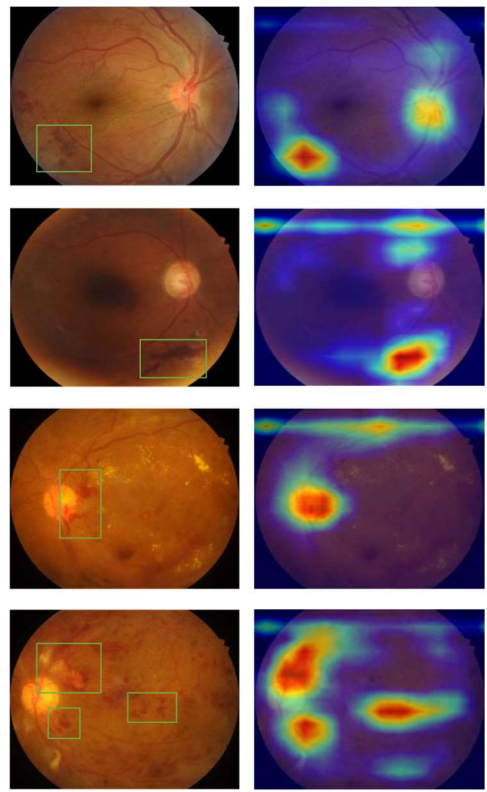

Fig. 5. The qualitative and attention visualization results of our proposed methods: the left column shows original images and the right column shows corresponding class activation maps. The green bounding boxes indicate hemorrhagic regions.

## V. Discussion

The grading performance of medical image small datasets is difficult to be improved due to the problems of insufficient data and class imbalance. In this work, we propose a multi-stage transfer learning method to fully learn the features of three datasets with different scales in order to solve the problem of insufficient data; furthermore, we introduce a CBCE loss into the DR classification task and retrain the FC layer with this loss only in the classifier learning stage to mitigate the class-imbalance problem. The experimental results on the IDRiD dataset demonstrate the effectiveness of our proposed methods.

Although our proposed methods have achieved a good performance, some limitations still exist. Firstly, the methods in this paper use classification models to extract the features of the whole image rather than incorporate pixel-wise lesion information and other prior knowledge to guide the classification. Secondly, this paper mainly focuses on the overall improvement of grading performance, especially that of serious classes such as DR3 and DR4. However, there may be cases where the performance of specific classes is not improved or even slightly decreased, such as mild and moderate. The reason for the above observations lies in two aspects: one is the mild class only accounts for 5% of the dataset, making it difficult to learn effective features; the other is mild and moderate are easily misclassified as adjacent classes due to high correlation between them (e.g., mild and No DR, moderate and mild are difficult to grade accurately even by experts). The solution to the above problems may further improve the grading performance of the small datasets. Thirdly, aggregated set from multiple datasets can be used for testing as a complement to IDRiD. The above issues will be taken into consideration and investigated in our future work.

## VI. Conclusion

In this paper, we apply multi-stage transfer learning and a class-balanced loss to improve the grading performance of small DR datasets. We decouple feature representation learning and classifier learning to design a new transfer learning framework. The feature representation stage trains on Eyepacs, DDR, and IDRiD datasets continuously to make the model learn representation features sufficiently. The classifier learning stage only trains the classifier with a class-balanced loss. The proposed scheme avoids searching for a complicated and time-consuming hyperparameter-tuning strategy and alleviates the problem of class imbalance in a simple way. The ablation studies on IDRiD dataset demonstrate the effectiveness of our proposed methods. Meanwhile, the proposed method outperforms several state-of-the-art methods as well. Our codes and weights will be available at GitHub.

## Acknowledgment

This work was partially supported by the National Natural Science Foundation of China (Grant No. 61261019), the Inner Mongolia Science & Technology Plan (Grant No. 201802027), and the Inner Mongolia Autonomous Region Natural Science Foundation (Grant No. 2018MS06023).

## References

[1] D.Ophthalmoscopy and E.Levels, "International clinical diabetic retinopathy disease severity scale detailed table," 2002.

[2] C.P.Wilkinson et al., "Proposed international clinical diabetic retinopathy and diabetic macular edema disease severity scales," Ophthalmology, vol. 110, pp.1677-1682, 2003.

[3] T.Li,Y.Gao,K.Wang, S.Guo, H.Liu, and H.Kang, "Diagnostic assessment of deep learning algorithms for diabetic retinopathy screening," Information Sciences, vol. 501, pp.511-522, 2019.

[4] Y.Zhou,X.He, S.Cui, F.Zhu, L.Liu, and L.Shao,"High-resolution diabetic retinopathy image synthesis manipulated by grading and lesions," International Conference on Medical Image Computing and Computer-Assisted Intervention, pp. 505-513,2019.

[5] Y.Zhou et al., "Collaborative learning of semi-supervised segmentation and classification for medical images," in Proceedings of the IEEE/CVF Conference on Computer Vision and Pattern Recognition, pp. 2079-2088, 2019.

[6] X.Li, X.Hu, L.Yu, L.Zhu, C.W.Fu, and P. A. Heng, "CANet: cross-disease attention network for joint diabetic retinopathy and diabetic macular edema grading," IEEE Transactions on Medical Imaging, vol.39, pp. 1483-1493,2019.

[7] X.He, Y.Zhou, B.Wang, S.Cui, and L.Shao, "Dme-net: Diabetic macular edema grading by auxiliary task learning," International Conference on Medical Image Computing and Computer-Assisted Intervention, pp.788-796, 2019.

[8] A.He, T.Li, N.Li, K.Wang, and H.Fu, "CABNet: Category attention block for imbalanced diabetic retinopathy grading," IEEE Transactions on Medical Imaging, vol.40, pp. 143-153, 2020.

[9] https://idrid.grand-challenge.org.

[10] B. Kang et al., "Decoupling representation and classifier for long-tailed recognition," arXiv preprint arXiv:1910.09217, 2019.

[11] B.Zhou, Q.Cui, X.S.Wei, and Z.M.Chen, "BBN: Bilateral-branch network with cumulative learning for long-tailed visual recognition," in Proceedings of the IEEE/CVF Conference on Computer Vision and Pattern Recognition, pp. 9719-9728,2020.

[12] Y.Cui, M.Jia, T.Y.Lin, Y.Song, and S.Belongie, "Class-balanced loss based on effective number of samples," in Proceedings of the IEEE/CVF Conference on Computer Vision and Pattern Recognition, pp. 9268-9277,2019.

[13] A.Galdran, J.Dolz, H.Chakor, H.Lombaert, and I.B.Ayed, "Cost-sensitive regularization for diabetic retinopathy grading from eye fundus images," in International Conference on Medical Image Computing and Computer-Assisted Intervention, pp.665-674,2020.

[14] S.Liu, L.Gong, K.Ma, and Y.Zheng, "GREEN: a graph residual re-ranking network for grading diabetic retinopathy," in International Conference on Medical Image Computing and Computer-Assisted Intervention, pp.585-594, 2020.

[15] https://www.kaggle.com/c/diabetic-retinopathy-detection/data.

[16] M.Tan and Q.Le, "Efficientnet: Rethinking model scaling for convolutional neural networks," International Conference on Machine Learning, pp.6105-6114, 2019.

[17] https://www.image-net.org/

[18] C.Huang, Y.Li, C. C.Loy, and X.Tang, " Learning deep representation for imbalanced classification," in Proceedings of the IEEE/CVF Conference on Computer Vision and Pattern Recognition, pp. 5375-5384, 2016.

[19] P.Porwal et al., "Idrid: Diabetic retinopathy–segmentation and grading challenge," Medical image analysis, vol.59, 2020.

[20] Z.Wu et al., "Coarse-to-fine classification for diabetic retinopathy grading using convolutional neural network," Artificial Intelligence in Medicine,vol.108,pp.1-9, 2020.

[21] N.S.Shaik and T.K.Cherukuri, "Lesion-aware attention with neural support vector machine for retinopathy diagnosis," Machine Vision and Applications, vol. 32, pp.1-13, 2021.

[22] T.Araújo et al.," DR| GRADUATE: Uncertainty-aware deep learning-based diabetic retinopathy grading in eye fundus images," Medical Image Analysis,vol.63, pp.1-17,2020.

[23] R.R.Selvaraju, M.Cogswell, A.Das, R.Vedantam, D.Parikh, and D.Batra, "Grad-cam: Visual explanations from deep networks via gradient-based localization,"in Proceedings of the IEEE International Conference on Computer Vision, pp. 618-626,2017.